\documentclass[journal=nalefd,manuscript=letter]{achemso}

\usepackage{bm}
\usepackage{multirow}
\usepackage{rotating}
\usepackage{mciteplus}
\usepackage{footmisc}

\author{Liujiang Zhou}
\affiliation{Bremen Center for Computational Materials Science, University of Bremen, Am Falturm 1, 28359 Bremen, Germany}
\alsoaffiliation{Max Planck Institute for Chemical Physics of Solids, Noethnitzer Str. 40, 01187 Dresden, Germany}
\email{liujiang86@gmail.com}
\author{Wujun Shi}
\affiliation{Max Planck Institute for Chemical Physics of Solids, Noethnitzer Str. 40, 01187 Dresden, Germany}
\alsoaffiliation{School of Physical Science and Technology, ShanghaiTech University, Shanghai 200031, China}
\author{Yan Sun}
\affiliation{Max Planck Institute for Chemical Physics of Solids, Noethnitzer Str. 40, 01187 Dresden, Germany}
\author{Bin Shao}
\affiliation{Bremen Center for Computational Materials Science, University of Bremen, Am Falturm 1, 28359 Bremen, Germany}
\author{Claudia Felser}
\affiliation{Max Planck Institute for Chemical Physics of Solids, Noethnitzer Str. 40, 01187 Dresden, Germany}
\author{Binghai Yan}
\affiliation{Max Planck Institute for Chemical Physics of Solids, Noethnitzer Str. 40, 01187 Dresden, Germany}
\alsoaffiliation{School of Physical Science and Technology, ShanghaiTech University, Shanghai 200031, China}
\author{Thomas Frauenheim}
\affiliation{Bremen Center for Computational Materials Science, University of Bremen, Am Falturm 1, 28359 Bremen, Germany}

\title[An \textsf{achemso} demo]
  {Rectangular Tantalum Carbide Halides TaCX (X = Cl, Br, I) monolayer: Novel Large-Gap Quantum Spin Hall Insulator}

\abbreviations{Quantum  spin  Hall  insulator, Tantalum Carbide Halide, Gapless edge states, Band inversion, First-principles calculations}

\begin{document}

\begin{abstract}
Quantum spin Hall (QSH) insulators possess edge states that are topologically protected from backscattering. However, known QSH materials (e.g. HgTe/CdTe and InAs/GaSb quantum wells) exhibit very small energy gap and only work at low temperature, hindering their applications for room temperature devices. Based  on  the  first-principles  calculations, we predict a novel family  of QSH insulators in monolayer tantalum carbide halide TaCX (X = Cl, Br, and I) with unique rectangular lattice and large direct energy gaps larger than 0.2 eV, accurately, 0.23$-$0.36 eV. The mechanism for 2D QSH effect in this system originates from a intrinsic \emph{d}$-$\emph{d} band inversion,  different from conventional QSH systems with band inversion between \emph{s}$-$\emph{p} or \emph{p}$-$\emph{p} orbitals. Further, stain and intrinsic electric field can  be used to tune the electronic structure and enhance the energy gap. TaCX nanoribbon, which has single-Dirac-cone edge states crossing the bulk band gap, exhibits a linear dispersion with a high Fermi velocity comparable to that of graphene. These 2D materials with considerable nontrivial gaps promise great application potential in the new generation of dissipationless electronics and spintronics.
\end{abstract}

Two-dimensional (2D) topological insulators (TIs), also known as Quantum spin Hall (QSH) insulators, characterized by an insulating bulk and fully spin-polarized gapless helical edge states without backscattering at the sample boundaries, are protected by time-reversal symmetry and thus are promising for achieving dissipationless transport devices.\cite{hasan_colloquium_2010, qi_topological_2011}  Compared to  three-dimensional (3D) TIs with the surface states only projected from exact 180$^{\circ}$C backscattering and suffering from other angles' scattering,  the
electrons with opposite spins in 2D TIs can only  move along two directions and  free from backscattering caused by nonmagnetic defects, thus holding  the unique advantages  and more promising application  for next-generation non-dissipative spintronic devices. A large bulk band gap is critical for the application of QSH insulators  in spintronic devices operating at room temperature. So far the already  realized 2D TIs exist in two quantum well systems, HgTe/CdTe\cite{bernevig2006d,konig_quantum_2007} and InAs/GaSb.\cite{Liu2008,knez_evidence_2011} However,  such  QSH insulators  exhibit energy gaps in the order of magnitude of 10 meV  and are usually  observed  in experiment only at ultra-high vacuum and extremely low temperature\cite{konig_quantum_2007,knez_evidence_2011}, hindering their application in realistic devices. Thus,
tremendous efforts have recently been performed to search for large-gap QSH materials, such as the prediction\cite{xu_large-gap_2013} and discovery \cite{zhu_epitaxial_2015}  of the stanene TI and similar materials~\cite{wu_prediction_2014,kou_robust_2014,song_quantum_2014,qian_quantum_2014,zhou_epitaxial_2014,li_giant_2015,luo_room_2015}.
It is worth being noted that most of the them  are those antimony (Sb)- or bismuth (Bi)-based compounds due to the extremely strong intrinsic spin-orbit coupling (SOC) strength of Bi/Sb atoms. Considering the importance of carbon (C) chemistry in our modern device applications, C-based QSH insulators~\cite{kane2005A} with sizable energy gaps are a  long-sought goal since the birth of TIs.

The existence of planar tetracoordinate carbon (ptC) (known as the anti-van't Hoff/Lebel\cite{siebert_compounds_1999} compound) in  molecules  and inorganic 2D nanosheets (such as 2D B$_2$C,\cite{wu_b2c_2009} Al-C systems\cite{dai_alxc_2014}) inspires the ptC-containing  2D materials' development. The  tetragonal titanium carbide (t-TiC) monolayer containing tetracoordinate-C atoms was formed by  strong d$-$p bonding interaction,  opening up a band gap up to 0.2 eV.\cite{zhang_two-dimensional_2012} In spite of its trivial band gap, the strong d$-$p bonding interaction between Ti and C atoms offers a new approach to realize C-based QSH insulator with large nontrivial gaps.  Moreover, impressive studies\cite{zhou_new_2015,liu_new_2015, sun_graphene-like_2015, qian_quantum_2014} show that transition-metal-based QSH insulators possess intrinsic and beyond \emph{s}$-$\emph{p} band inversion, such as \emph{p}$-$\emph{p}\cite{weng_transition-metal_2014}, \emph{d}$-$\emph{p}\cite{qian_quantum_2014} as well as \emph{d}$-$\emph{d}\cite{zhou_new_2015}, due to the strong electronic interaction instead of SOC.\cite {yang_d-p_2014} Such intrinsic band inversion not originating from SOC greatly enrich the family of QSH insulators and enlarge their promise for nanoscale device applications, and also stimulate the further investigations on interesting phenomena, such as transition in correlated Dirac fermions\cite{yu_mott_2011} and interaction induced topological Fermi liquid.\cite{castro_topological_2011}
It is noteworthy that transition-metal-based 2D materials usually holding the intrinsic layered characteristic, such as transition metal-based dichalcogenides (such as MoS$_2$, TiS$_2$, TaS$_2$),\cite{xu_graphene-like_2013} transition-metal carbides (MXenes),\cite{bhimanapati_recent_2015} transition-metal nitrides (such as TiNCl)\cite{liu_single-layer_2014} as well as transition-metal Halides (e.g., ZrBr),\cite{zhou_new_2015} can be achieved from corresponding 2D intrinsic layered materials via simple exfoliation or chemical vapor deposition,  without any chemical functionalization or lattice distortion or applying strain, which is beneficial for the future experimental preparation for  monolayers in the field of QSH insulators. A nature question then arises: Does QSH effect exist in transition-metal carbide-based 2D compounds not containing Sb/Bi atoms and possessing high feasibility in experiment? Addressing this question successfully will not only enrich our physics of 2D TIs but also pave new ways for designing C-based topological materials for realistic applications.

Here, we predict a new family of  QSH insulators based on monolayer  tantalum carbide halide TaCX (X=Cl, Br, I)  containing quasiplanar tetracoordinate C atoms. All these TaCX monolayers are robust QSH insulators against external strain, showing very large tunable band gaps  in the range of 0.23$-$0.36 eV, comparable with  recent synthesized stanene (0.3 eV)\cite{zhu_epitaxial_2015} and larger than monolayer HfTe$_5$ (0.1 eV)\cite{weng_transition-metal_2014} and recent sandwiched graphene-based heterostructures (30$-$70 meV).\cite{kou_graphene-based_2013, kou_robust_2014} The phonon spectrum calculations further suggest that the freestanding monolayer structure can be stable. Interestingly, TaCX monolayer possesses a novel band inversion between the \emph{d}$-$\emph{d} orbitals, distinctive from conventional TIs, greatly enriching the family of QSH insulators.  Additionally, spin-orbit coupling opens a gap that is tunable by external electric field and strain.

\section{Results and Discussion}

The ternary tantalum  carbide halides (TaCX, X = Cl, Br, I) possess the FeOCl structure type, as observed in transition metal nitride halides (such as $\alpha$-TiNCl monolayer\cite{liu_single-layer_2014}), consisting of buckled double Ta$-$C layers sandwiched between halogen atomic layers. They crystallize in the orthorhombic layered structure with space group \emph{Pmmn},  showing the $D_{2h}$ symmetry. Figure 1a$-$c present the orthorhombic atomic structure of the TaCX  films in their most stable configuration, in which each of C atoms is quasiplanar tetracodinated with ambient Ta atoms and each of Ta atoms is six-coordinated with ambient four C atoms and two halogen atoms. The optimized lattice constants are listed in Table 1. The lattice constant $a$ has an increase trend from 3.41 to 3.63 $\mathrm{\AA}$ as the increasing of atomic radius of X atom in TaCX monolayer. While, the lattice constant $b$ almost maintain the constant value of about 4.23 $\mathrm{\AA}$. The stability of the TaCX sheet can  be understood by analyzing its deformation electron density, which reveals electron transfer from the Ta to C atom as shown in Figure 1e. The transferred electrons are mainly from the Ta-d$_{z^2}$ state, delocalized around Ta$-$C bonds. Meanwhile, the C-p$_{z}$ is also found to partially deplete and delocalized over the Ta$-$C bonds. Such a delocalization between d$_{z^2}$ and p$_{z}$ is crucial to stabilizing the quasiplanar
tetracoordinate atoms, because it not only weakens atomic activity in forming out-of-plane bonds but also strengthens in-plane Ta-C bonds, in analogy to that in t-TiC monolayer.\cite{zhang_two-dimensional_2012} The role of halogen atom X is used to saturate the  dangling bonds from remaining d-orbitals of Ta atom due to the fact that Nb atom holds one more d electron than Ti atom in  t-TiC monolayer. The calculated phonon spectrums(Figure S1) show  no negative frequency for monolayer TaCX, which suggests it is a stable phase without any dynamical instability, related to the interactions between atomic layers.

The  electronic  band  structures  of  monolayer TaCX  calculated based on PBE  level are shown in Figure 2a$-$c.  In the absence of SOC, all these TaCX monolayers show semimetal feature with the appearance of two Dirac cones  centered at finite momenta on $\mathrm{Y}-\Gamma-\mathrm{Y}$ in 2D Brillouin zone (BZ) (see in Figure 1d) due to the band inversion. When the  SOC is included, the degeneracy of the Dirac point is lifted out, and then the energy gaps of 0.12, 0.16 and 0.26 eV are opened up at the Dirac points for TaCCl, TaCBr and TaCI monolayers, respectively. The conduction and valence bands display a camelback shape near $\Gamma$ in  BZ, suggestive of the band inversion with a large inverted gap (2$\delta$) at $\Gamma$ of about 1.52 eV, located at $\Lambda$ = $\pm$(0,0.168)$\mathrm{\AA}^{-1}$ (red dots in Figure 1d). The fundamental gap ($E_\mathrm{g}$) and inverted gap (2$\delta$) of all TaCX are shown in Figure 2 and listed in table 1.  Because the TaCX structure has inversion symmetry, we have investigated the $Z_2$ topological invariant \cite{kane_$z_2$_2005,kane_quantum_2005} by evaluating the parity eigenvalues of occupied states at four time-reversal-invariant-momentum(TRIM) points in BZ.\cite{fu_topological_2007} Although the products of the parity eigenvalue at the $\Gamma$ can be calculated to be $-$1,  the parity eigenvalues at the other three TRIM points can not be accessible  in the same way due to the double-degenerated band feature below Fermi level (see in Figure 2a$-$c), which leads
to  each of the products of  parity  at such TRIM points to be +1.  It implies that the QSH effect can be realized with Z$_2$ topological invariant $\nu_0$ = 1 in  TaCX monolayer. Such a nontrivial topological nature can also be confirmed by the evolution of the Wannier centers (see calculation detail in Methods). As shown in Figure 2d$-$f, the evolution lines of Wannier centers cross the arbitrary reference line an odd number of times in the $k_z$ = 0 plane, indicating TaCX monolayer is the QSH insulator. Since the PBE functional is known to usually underestimate the band gap, we have performed additional calculations for TaCX monolayer using hybrid functional (HSE06)\cite{heyd_erratum,heyd_hybrid_2003} to correct the band gaps. The nontrivial gap of TaCX has an enlargement by about 0.1 eV. Interestingly, The tunable HSE06 band gaps in the range of $0.23-0.36$ eV (Table 1), are comparable with stanene (0.3 eV) \cite{xu_large-gap_2013} and larger than HfTe$_5$ (0.1 eV) monolayer \cite{weng_transition-metal_2014} and  sandwiched graphene-based heterstructures (30$-$70 meV).\cite{kou_graphene-based_2013, kou_robust_2014} The comparatively large tunable nontrivial gaps in a pure monolayer materials without chemical doping, or field effects, which are very beneficial for the future experimental preparation for monolayer TaCX  and makes them highly adaptable in various application environments.

To illustrate the band inversion process explicitly, atomic orbitals  \emph{d}$_{xy}$, \emph{d}$_{yz}$ of Ta at the  $\Gamma$ point around the Fermi level  are present in Figure 3a for TaCI under different hydrostatic strains. In the chemical bonding process, Ta atoms move to each other,  Ta-\emph{d}$_{xy}$  states with parity $p$ = +1  are pulled down, while the Ta-\emph{d}$_{yz}$ orbitals with $p = -1$ are shifted into the conduction bands, leading to a level crossing with the parity exchange between occupied and unoccupied bands. Such a cross therefore induces a topological phase transition from a trivial insulator to a TI between \emph{d}$_{xy}$  and   \emph{d}$_{yz}$ orbitals of Ta atoms at the critical stretching lattice of about $198\%$ $a$. It is noteworthy that the \emph{d}$-$\emph{d} band inversion is distinctive from conventional band inversion from \emph{s}$-$\emph{p} orbitals, or \emph{p}$-$\emph{p} orbitals. When including the effect of SOC on equilibrium structure,  their parities remain unchanged, which is also confirmed by the same $Z_2$ invariant of TaCX monolayer (Table 1). Thus, the band inversion does not originate from the SOC. The role of SOC is only to open up a fundamental gap, which is similar to  HfTe$_5$\cite{weng_transition-metal_2014,wu_experimental_2016} and square-octagonal MX$_2$ structures, \cite{sun_graphene-like_2015, ma_quantum_2015} and 1T'-MoX$_2$\cite{qian_quantum_2014} and ZrBr family,\cite{zhou_new_2015} through a similar mechanism as Kane-Mele model for graphene.\cite{kane_quantum_2005}

As we can see the strain-induced band inversion in TaCX, it is quite crucial to check the robustness of topological
nature against external strain. We impose biaxial strain on the 2D planes of these systems by turning the planar lattice parameter.  The magnitude of strain is described by $\epsilon  = \Delta{a}/a_0$. Here, $a_0$ and $a = \Delta{a} + a_0$ denotes the lattice parameters of the unstrained and strained systems, respectively. The products of the parity eigenvalues at the four TRIM points and the feature of evolution of Wannier centers are not destroyed in the range of $-8\%$ to 6\%,  $-8\%$ to 6\% and  $-8\%$ to 6\% for  TaCCl, TaCBr and TaCI, respectively, suggesting the nontrivial topological phase are indeed insensitive to the applied strain.  Moreover, as shown in Figure 3b, the magnitude of nontrivial global bulk band gap can be modified significantly via the interatomic coupling subjected to external strain field, offering an effective way to tune the topological properties of these new 2D TIs and benefiting the potential application in spintronics.

The QSH insulator phase in monolayer TaCX  should support an odd number of topologically protected gapless conducting edge states connecting the valence and conduction bands of each system at certain $k$-points, protected from localization and elastic backscattering by time-reversal symmetry. Figure 4 displays the edge states of TaCX obtained from surface
Green's function calculations plus Wannier functions\cite{sancho_highly_1985} extracted from $ab$ $initio$ calculations (see Methods). The Dirac cones are all located at the $\Gamma$ point for three cases. At a given edge, two counter-propagating edge states carry opposite spin-polarizations, a typical feature of the 1D helical state of a QSH phase. The pair of counter-propagating edge modes inside the bulk gap possess a high velocity of $\sim1.5\times 10^5$ $m/s $, comparable to that of graphene, which is  beneficial to future dissipationless transport. Although the details depend on different compounds and edges, the non-trivial $Z_2$ invariant and evolution of Wannier centers guarantee the edge bands always cutting Fermi level odd times.

The inverted bands between  Ta-d orbitals within unwell-separated two atomic planes sandwiched between two X atom layers offer a facile approach to modulate the topological band gap. For instance, by substituting the halogen atom (X) by another halogen atom (X') in one of outer halogen layers so as to generate a non-centro-symmetric structure TaCXX', the in-phase alignment of the dipoles within X and X' atom layers along z axis can be achieved,  equivalent to applying an intrinsic electric field perpendicular to the $xy$ plane. The electronic structure of TaCXX' shows that the nontrivial band gap has a  tunability by 30-70 meV (Figure S2 and Table 1). The nontrivial topological feature  still exist in these three TaCXX' monolayers, guaranteed by the evolution of Wannier certers (Figure S3) and topologically protected edge states (Figure S4), further confirming the robustness of topology in TaCXX' subjected to an intrinsic electric field.

In summary,   monolayer TaCX (X = Cl, Br, and I) containing tetracoordinate carbon atoms constitutes a novel family of robust QSH insulators in 2D transition metal carbide halides  with large tunable nontrivial gaps larger than 0.23 eV. The mechanism for the QSH effect originates from the band inversion between \emph{d}$-$\emph{d} orbitals of Ta atoms, different from conventional band inversion from \emph{s}$-$\emph{p}, or  \emph{p}$-$\emph{p} orbitals. Stain and intrinsic electric field can  be used to tune the electronic structure and enhance the energy gap. TaCX nanoribbon has single-Dirac-cone edge states crossing the bulk band gap, exhibiting a linear dispersions with a high Fermi velocity comparable to that of graphene. These comparable large nontrivial gaps in pure monolayer materials without external strain, or distortion, which are very beneficial for the future experimental preparation, makes them highly adaptable in various application environments. These interesting results may
stimulate further efforts on carbon chemistry  and transition-metal-based QSH insulators.

\section{Methods}

Ground-state atomic structures of all Tantalum Carbide Halides TaCX were fully relaxed using first-principles calculations  density functional theory (DFT) implemented within the Vienna $ab$ $initio $ Simulation Package (VASP) \cite{kresse_efficient_1996}. The exchange correlation interaction is treated within the generalized gradient approximation(GGA) \cite{perdew_generalized_1996}, which is parametrized by the Perdew, Burke, and Ernzerhof (PBE)\cite{perdew_self-interaction_1981}. An energy cutoff of 500 eV  was used in all calculations. All the atoms in the unit cell were fully relaxed until the maximum residual force force on each atom was less than 0.01 eV/{\AA}. In the structural relaxation and the stationary self-consistent-field calculation,  $k$-point grids with 0.025 and 0.015 $\mathrm{\AA}^{-1}$ spacing were adopted, respectively. A large vacuum region of more than 15 \AA was applied to the plane normal direction in order to minimize image interactions from the periodic boundary condition. Since DFT methods often underestimate the band gap, the screened exchange hybrid density functional by Heyd-Scuseria-Ernzerhof (HSE06) \cite{heyd_erratum,heyd_hybrid_2003} is adopted to correct the PBE band gaps. The phonon dispersion calculations were calculated by Phonopy code\cite{baroni_phonons_2001} based on a supercell approach.

The topological character was calculated according to the Z$_2$ invariant by explicitly calculating band parity of the materials with inversion symmetry, outlined in Ref. \cite{fu_topological_2007}. Regarding the topological trait of TaCXX' without inversion center,  the Z$_2$ invariant was computed by tracing the Wannier charge centers using the non-Abelian Berry connection.\cite{soluyanov_computing_2011,yu_equivalent_2011} To reveal the helical edge states of monolayer TaCX explicitly,  we performed iterative Green¡¯s function calculations\cite{sancho_highly_1985} using tight-binding Hamiltonian  by projecting the Bloch states into Wannier functions.\cite{mostofi_wannier90:_2008, franchini_maximally_2012} The imaginary part of the surface Green's function is relative to the local density of states (LDOS), from which we can obtain the edge states.\cite{shi_converting_2015}

\clearpage
\newpage
\begin{suppinfo}
The phonon spectrums of TaCX monolayer, the band structures, evolution lines of Wannier centers and edge states for TaCXX' monolayer. This material is available free of charge via the Internet at http://pubs.acs.org
\end{suppinfo}

\section{Notes}
The authors declare no competing financial interests.

\begin{acknowledgement}
L.Z and B.S acknowledges financial support from Bremen University. B.Y. and C.F. acknowledge financial support from the European
Research Council Advanced Grant (ERC 291472). The support of the Supercomputer Center of Northern Germany (HLRN Grant No. hbp00027).
\end{acknowledgement}

\clearpage
\newpage

\begin{figure*}
\centering
\includegraphics[width=12cm]{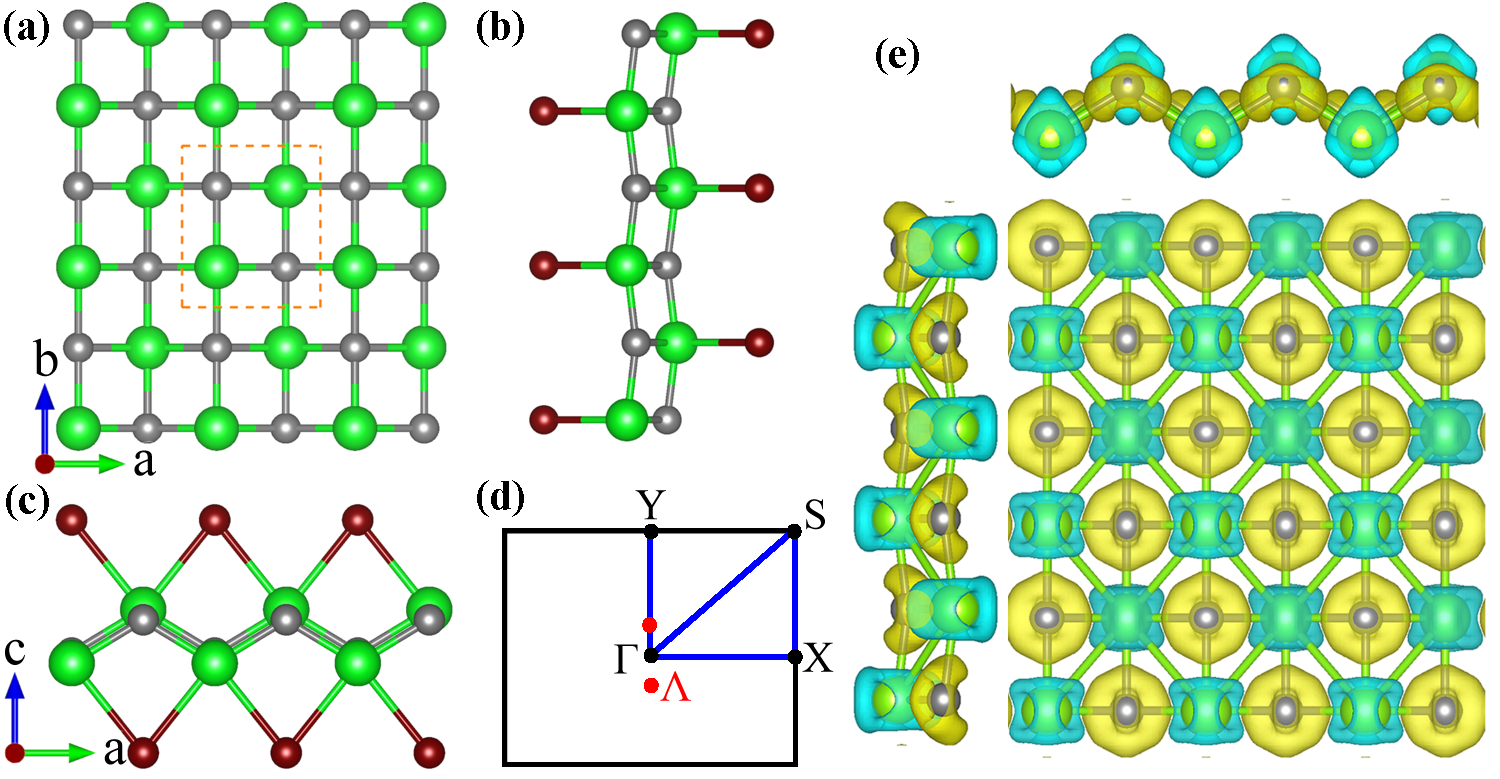}\\
\caption{The crystal structures of ternary transition-metal carbide halides TaCX (X = Cl, Br, I) in the form of  FeOCl type:(a) the top view; (b,c) the side view along a and b axis, respectively.  The unit cell is indicated by red dashed lines. (d) The Brillouin zone (BZ) of 2D TaCX. The locations of the fundamental gap are marked by red dots and labeled by $\Lambda$. (e) Isosurface plots (0.015 e/\AA) of deformation electronic density. Charge accumulation and depletion regions are shown in yellow and blue, respectively. For the sake of clarity, halogen atoms are omitted. }
\label{Fig. 1}
\end{figure*}

\newpage
\begin{figure}
\centering
\includegraphics[width=10cm]{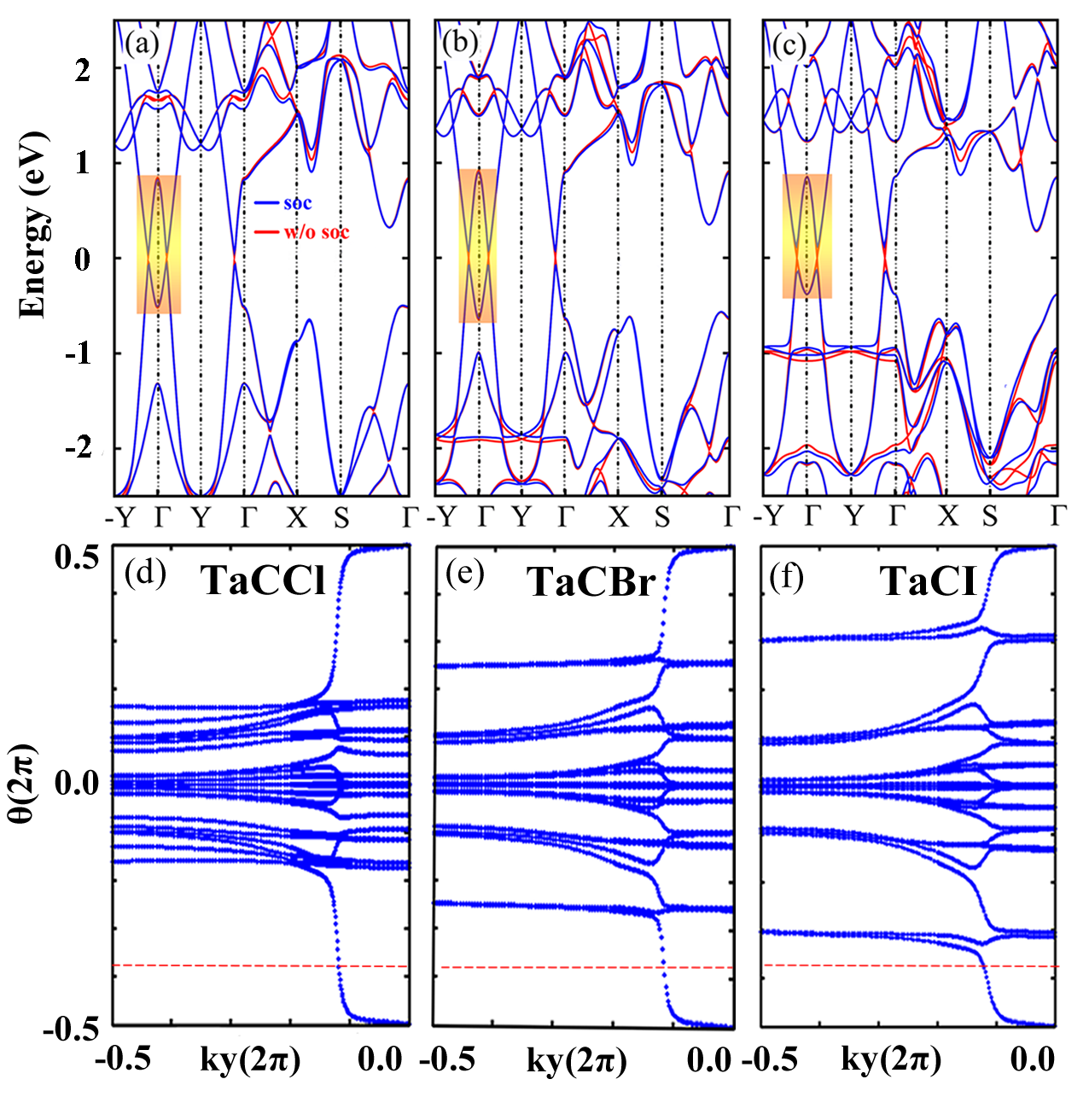}
\caption{Electronic band structures of (a) TaCCl, (b) TaCBr, and (c) TaCI. The red and bule lines correspond to band structures without and with SOC, respectively. The nontrivial band gaps are  indicated. The camelback shape dispersion from  conduction and valence bands near $\Gamma$ in the 2D Brillouin zone (BZ) are indicated in marked square.  The large inverted gaps (2$\delta$) at $\Gamma$ for  TaCCl,  TaCBr, and TaCI  are 1.52, 1.54 and 1.50 ev, respectively. The Fermi energy is set to 0 eV. The evolution lines of Wannier centers for (d) TaCCl, (e) TaCBr, (f) TaCI monolayers. For the three systems, the evolution lines cross the reference line an odd number of times in the k$_z$ = 0 plane, indicating such three 2D monolayers are the QSH insulators.}
\label{Fig. 2}
\end{figure}
\newpage

\begin{figure*}
\centering
\includegraphics [width=12cm]{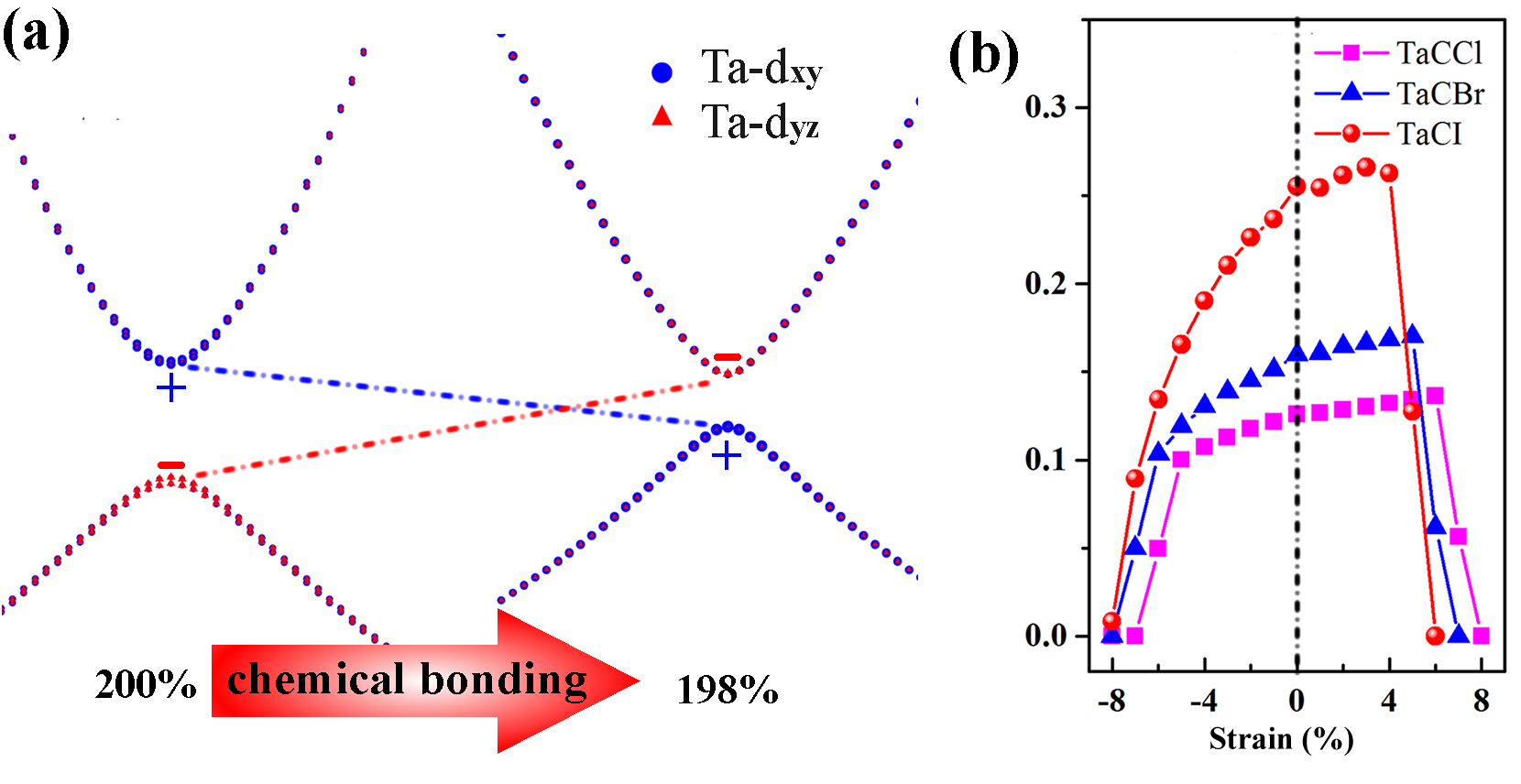}\\
\caption{(a) Illustration of the evolution from the atomic orbitals \emph{d}$_{xy}$, \emph{d}$_{yz}$ of Ta at the  $\Gamma$ point around the Fermi level under different hydrostatic strains in the chemical bonding process.  Parity values are presented near the various orbitals.  The band inversion will occur between  \emph{d}$_{xy}$ and \emph{d}$_{yz}$ orbitals when stretching lattice parameter \emph{a} with the lattice symmetry preserved. (b) Strain dependencies of the global bulk band gap of  TaCCl,  TaCBr and  TaCI monolayers with SOC. }
\label{Fig. 3}
\end{figure*}

\newpage

\begin{figure}
\centering
\includegraphics[width=8cm]{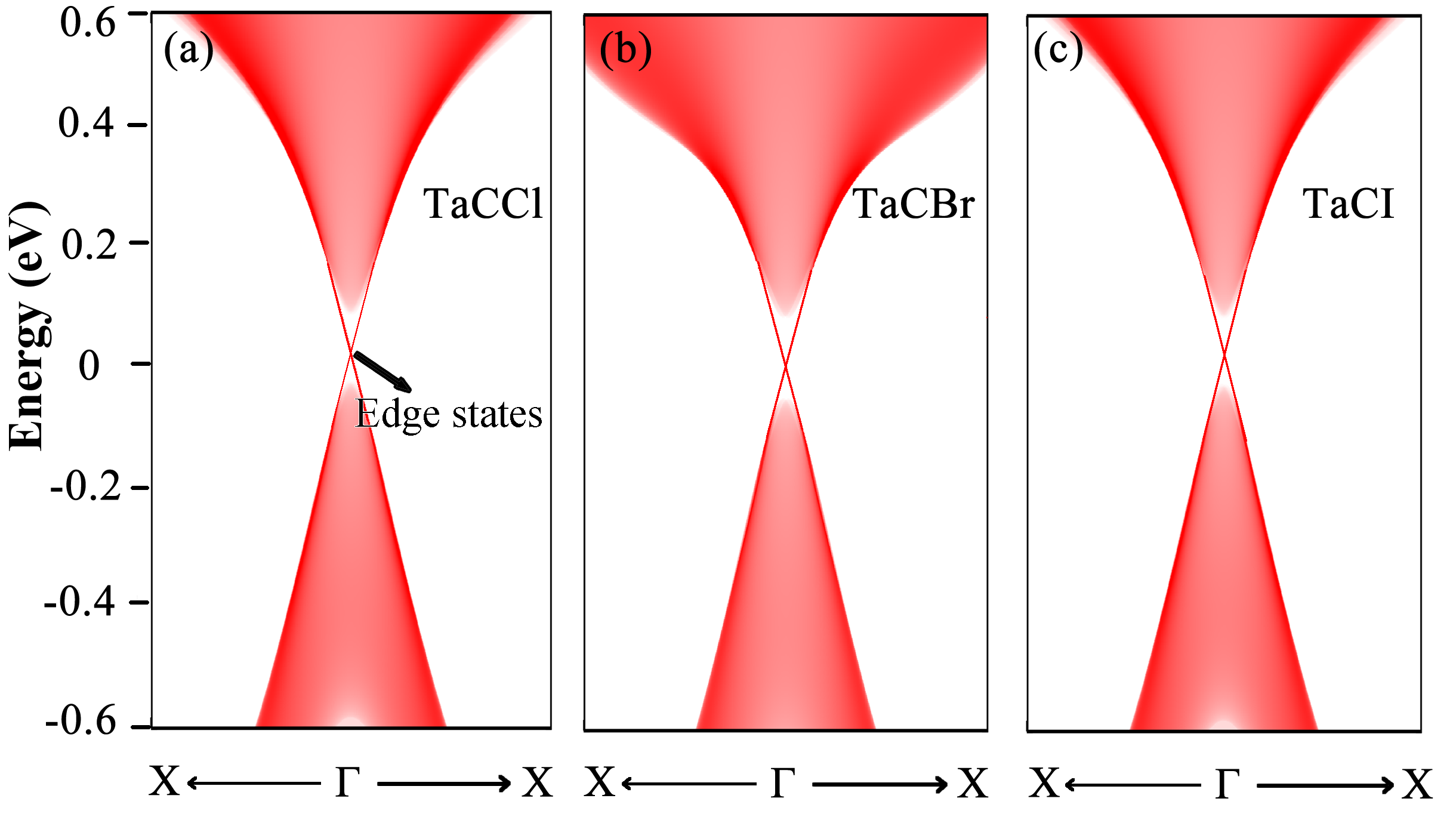}\\
\caption{The calculated topological edge states of  (a) TaCCl, (b) TaCBr and (c) TaCI monolayers with SOC obtained from the tight-binding Wannier function method. The Fermi energy is set to 0 eV. Clearly, Dirac surface states emerge in the bulk gap. }
\label{Fig. 4}
\end{figure}

\newpage

\begin{table}
\caption{\label{tab:1} The predicted lattice constants of monolayer TaCX, and their band gaps (eV) with SOC based on PBE and HSE06.}
\begin{center}
\begin{tabular}{l c c c c c  c}
\hline
 Compound &  TaCCl & TaCBr & TaCI& TaCClBr & TaCICl & TaCIBr \\
\hline
   Lattice a ({\AA})     & 3.41	     & 3.49	 & 3.63	   & 3.45	 & 3.52 & 3.56       \\
   Lattice b ({\AA})     & 4.23       & 4.23  & 4.24    & 4.229   &4.225  &   4.234  \\
   PBE gap               & 0.13	     & 0.16	& 0.26	   & 0.14	& 0.19	 & 0.21 \\
   HSE$06$ gap           & 0.23	     & 0.26	& 0.36	   & 0.30	   & 0.29	& 0.31  \\
   $Z_2$  invariant ${\nu_0}$   & 1          & 1       & 1        & 1      &1     &   1    \\
\hline
\end{tabular}
\end{center}
\end{table}
\newpage
\bibliography{TaCX}

\begin{tocentry}
\includegraphics [width=6.0cm] {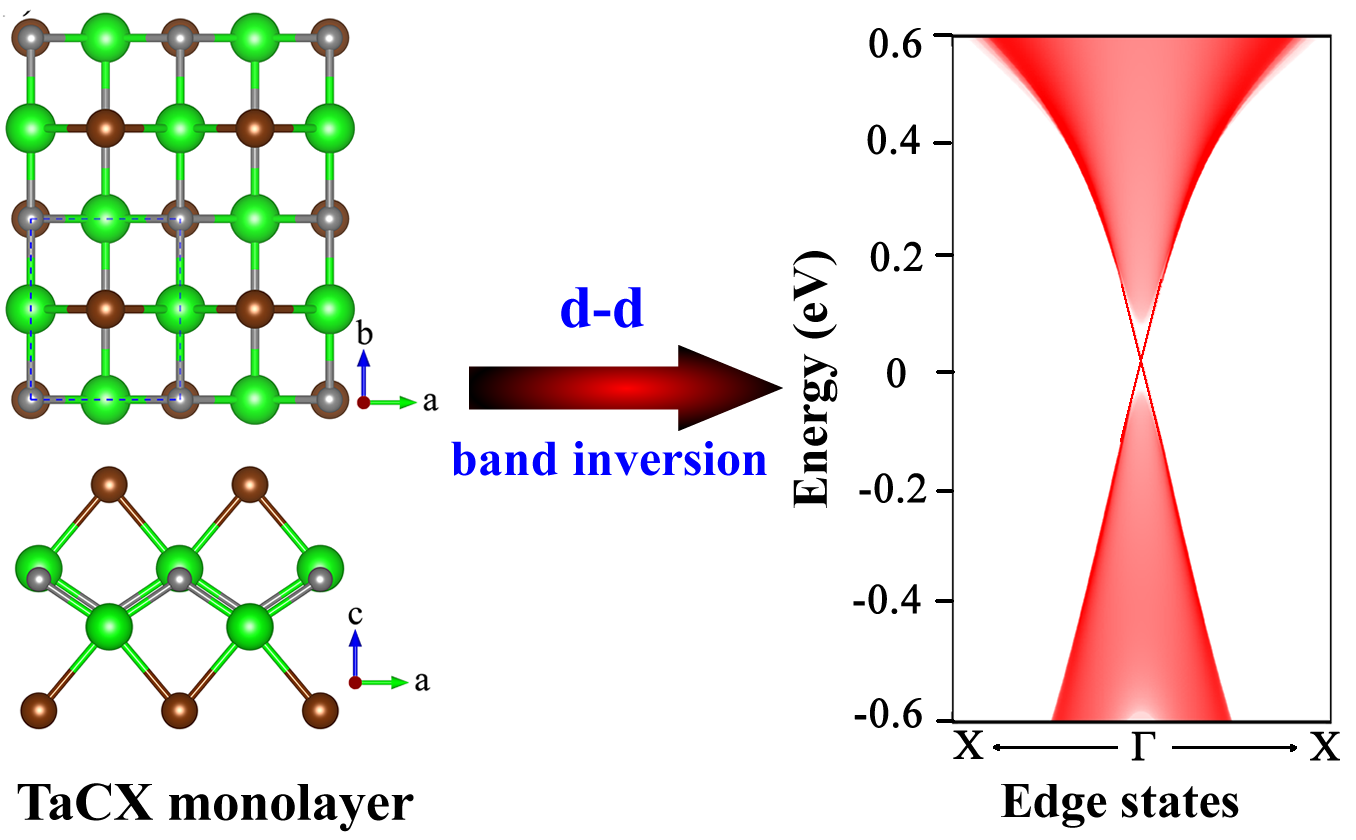}

 Based  on  the  first-principles  calculations, we predict a novel family  of two-dimensional (2D) QSH materials  in monolayer tantalum carbide halides  TaCX (X=Cl, Br, and I) with large nontrivial gaps of 0.23$-$0.36 eV. A novel $d$-$d$ band inversion is responsible for the 2D QSH effect, distinctive from conventional band inversion between $s$-$p$ orbitals, or $p$-$p$ orbitals. Stain and intrinsics electric field can  be used to tune the electronic structure and enhance the energy gap. TaCX nanoribbon, which has single-Dirac-cone edge states crossing the bulk band gap, exhibits a linear dispersions with a high Fermi velocity comparable to that of graphene.
\end{tocentry}

\end{document}